\documentclass[twoside]{article}
\usepackage[dvips]{graphicx}
\usepackage{qic,epsfig}
\usepackage{bm}
\usepackage{amsmath,amssymb}
\usepackage{type1cm}

\newtheorem{theo}{Theorem}
\renewcommand{\thefootnote}{\fnsymbol{footnote}}  
\DeclareSymbolFont{lettersA}{U}{txmia}{m}{it}
 \DeclareMathSymbol{\Indi}{\mathord}{lettersA}{'211}
 \usepackage{latexsym}
\begin{document}
\setlength{\textheight}{8.0truein}    
          
\normalsize\textlineskip
\thispagestyle{empty}
\setcounter{page}{1}

\vspace*{0.88truein}

\alphfootnote

\fpage{1}
\centerline{\bf Weak limit theorem of a two-phase quantum walk with one defect}
\vspace*{0.37truein}
\centerline{\footnotesize
Shimpei Endo\footnote{shimpei.endo@lkb.ens.fr}}
\vspace*{0.015truein}
\centerline{\footnotesize\it Laboratoire Kastler Brossel, Ecole Normale Superieure,}
\baselineskip=10pt
\centerline{\footnotesize\it  24 rue Lhomond, 75231 Paris, France}
\vspace*{10pt}
\centerline{\footnotesize
Takako Endo\footnote{g1170615@edu.cc.ocha.ac.jp\;(e-mail of the corresponding author)}}
\vspace*{0.015truein}
\centerline{\footnotesize\it Department of Physics, Ochanomizu University}
\baselineskip=10pt
\centerline{\footnotesize\it 2-1-1 Ohtsuka, Bunkyo, Tokyo, 112-0012,
Japan}
\vspace*{10pt}
\centerline{\footnotesize 
Norio Konno\footnote{konno@ynu.ac.jp}}
\vspace*{0.015truein}
\centerline{\footnotesize\it Department of Appllied Mathematics, Faculty of Engineering, Yokohama National University}
\baselineskip=10pt
\centerline{\footnotesize\it 79-5 Tokiwadai, Hodogaya, Yokohama, 240-8501, Japan}
\vspace*{10pt}
\centerline{\footnotesize 
Etsuo Segawa\footnote{e-segawa@m.tohoku.ac.jp}}
\vspace*{0.015truein}
\centerline{\footnotesize\it Graduate School of Information Sciences, Tohoku University}
\baselineskip=10pt
\centerline{\footnotesize\it 6-3-09 Aramaki Aza, Aoba, Sendai, Miyagi, 980-8579, Japan}
\vspace*{10pt}
\centerline{\footnotesize 
Masato Takei\footnote{takei@ynu.ac.jp}}
\vspace*{0.015truein}
\centerline{\footnotesize\it Department of Appllied Mathematics, Faculty of Engineering, Yokohama National University}
\baselineskip=10pt
\centerline{\footnotesize\it 79-5 Tokiwadai, Hodogaya, Yokohama, 240-8501, Japan}
\vspace*{10pt}

\vspace*{0.225truein}
\vspace*{0.21truein}
\begin{abstract}

We attempt to analyze a one-dimensional space-inhomogeneous quantum walk (QW) with one defect at the origin, which has two different quantum coins in positive and negative parts. We call the QW ``the two-phase QW", which we treated concerning localization theorems \cite{endosan}. The two-phase QW has been expected to be a mathematical model of the topological insulator \cite{kitagawa} which is an intense issue both theoretically and experimentally \cite{be,chen, fu}.
In this paper, we derive the weak limit theorem describing the ballistic spreading, and as a result, we obtain the mathematical expression of the whole picture of the asymptotic behavior.
Our approach is based mainly on the generating function of the weight of the passages. 
We emphasize that the time-averaged limit measure is symmetric for the origin \cite{endosan}, however, the weak limit measure is asymmetric, which implies that the weak limit theorem represents the asymmetry of the probability distribution.

\end{abstract}

\vspace*{1pt}\textlineskip 
\section{Introduction} 
This paper is a sequential work of \cite{endosan}.
For its characteristic properties, quantum walks (QWs) have attracted much attention of various fields, such as, quantum search algorithms \cite{ambainis,kempe}, and topological insulators \cite{kitagawa}. Owing to such applications of the QWs, it is of great importance to study the QWs both analytically and numerically, and indeed,
many researchers have tried to investigate the asymptotic behaviors of QWs from various viewpoints \cite{cho,joye,segawa,konnoweak,yoo,wojcik} in the past decade. 
From the mathematical view points, two kinds of limit theorems for QWs have been constructed so far.
The one is localization theorem. 
Localization is one of the typical properties of discrete-time QWs, which was first studied by Inui {\it et al.} \cite{inui} both mathematically and numerically. 
The detailed definition of localization is found in \cite{scholz,joye} for example. 
The other is the weak limit theorem whose typical expression is described as follows \cite{segawa}:
There exist $C\in[0,1),\;a\in(0,1)$, and a rational polynomial $w(x)$ such that 
\begin{align}\mu(dx)=C\delta_{0}(dx)+w(x)f_{K}(x;a)dx\label{weakmeasure}\end{align}
where
\begin{align}f_{K}(x;a)=\frac{\sqrt{1-a^{2}}}{\pi(1-x^{2})\sqrt{a^{2}-x^{2}}}I_{(-a,a)}(x)\label{konnofunction}\end{align}
with
\begin{eqnarray*}I_{A}(x)=\left\{ \begin{array}{ll}
1&(x\in A)\\
0&(x\notin A)
\end{array} \right..
\end{eqnarray*} 
We should note that the first term, Dirac measure part in Eq. \eqref{weakmeasure}, $C\delta_{0}(dx),$ corresponds to localization, and the second term, absolutely continuous part, $w(x)f_{K}(x;a)dx$, corresponds to the ballistic spreading. We remark that Eq. \eqref{weakmeasure} gives
\[1=C+\int^{\infty}_{-\infty}w(x)f_{K}(x;a)dx.\]
\indent
So far, the weak limit theorem of one-dimensional space-homogeneous QWs, such as Hadamard walk \cite{konnoweak}, Grover walk \cite{kota}, have been derived. 
In $2013$, Konno {\it et al.} \cite{segawa} have first given the weak limit theorem for the typical inhomogeneous QWs, taking advantage of the generating function of the weights of passages. The method permits the analysis only for the QWs with one defect at the origin, whose quantum coins are the same both in positive and negative parts.
Recently, various kinds of methods have been constructed to investigate mathematically the asymptotic behavior of QWs, such as
the Fourier analysis \cite{schudo}, the CGMV method \cite{cantero}, the stationary phase method \cite{nayak}, the path counting method \cite{konnopath}, and the generating function method \cite{endo}. 
We can expect to analyze various kinds of inhomogeneous QWs by the generating function method, while the Fourier analysis and stationary phase method are useful to study homogeneous QWs. 
However, it has not been clear the types of QWs that can be analyzed by the generating function method. 
We can also analyze inhomogeneous QWs via the CGMV method, still the CGMV method allows only for the general discussion of localization properties for the typical QWs in one dimension. 
The generating function method offers not only localization theorem, but also the weak limit theorem for QWs.\\
\indent
By using the generating function method, we focus on the ballistic behavior of ``the two-phase QW''.  It has been known that the two-phase QW is deeply related to the topological insurator which has attracted much attention recently of many physicists \cite{be,fu, kitagawa}. Hence we expect that  the two-phase QW can be utilized to study the topological insurator as its mathematical model. Therefore it would be greatly worth to study the mathematical aspects of the two-phase QW to exactly grasp the asymptotic behavior.
Our main result is the first application of the generating function method to the weak limit theorem of the two-phase QW.
Combaining the time-averaged limit measure \cite{endosan} with the result in this paper, we obtain the whole mathematical picture of the asymptotic behavior of our two-phase QW.\\
\indent
The rest of this paper is organized as follows. In Section $2$,
we define the two-phase QW which is the main target in this paper, and present our main result. 
In Section $3$, we give the proof of Theorem $1$.

 
\section{Model and the results}

\subsection{The two-phase QW}
For the general setting of discrete-time QW in one dimension, the walker has a coin state at position $x$ in each time $t$ described by a two-dimensional vector as follows:
\[ \Psi_{t}(x)=
\begin{bmatrix}
\alpha_{t}(x) \\ 
\beta_{t}(x)
\end{bmatrix} 
\;\;\;(x\in \mathbb{Z},\;\alpha_{t}(x),\beta_{t}(x)\in \mathbb{C}),\]
where $\mathbb{C}$ is the set of complex numbers.

In this paper, we focus on a discrete-time QW with
two phases in one dimension defined by the unitary matrices as follows:
\begin{align}
U_{x}=\left\{ \begin{array}{ll}
U_{+}=\dfrac{1}{\sqrt{2}}\begin{bmatrix}
1 & e^{i\sigma_{+}} \\
e^{-i\sigma_{+}} & -1 \\
\end{bmatrix}&(x\geq 1),\\
\\
U_{-}=\dfrac{1}{\sqrt{2}}\begin{bmatrix}
1 & e^{i\sigma_{-}} \\
e^{-i\sigma_{-}} & -1\\
\end{bmatrix}&(x\leq -1),\\
\\
U_{0}=\begin{bmatrix}
1 & 0 \\
0 & -1 \\
\end{bmatrix}& (x=0),
\end{array} \right.\label{2phase_def}
\end{align}
where $\sigma_{\pm}\in[0,2\pi)$.
The time evolution is determined by the recurrence formula
\begin{align*}
\Psi_{t+1} (x) = P_{x+1} \Psi_t (x+1) + Q_{x-1} \Psi_t (x-1) \quad (x \in \mathbb{Z}),
\end{align*}
where
\begin{align*}
P_x =\left\{ \begin{array}{ll}
\dfrac{1}{\sqrt{2}}\begin{bmatrix} 
1 & e^{i\sigma_{+}} \\ 
0 & 0 
\end{bmatrix}& (x\geq1), \\
\\
\begin{bmatrix} 
1 & 0 \\ 
0 & 0 
\end{bmatrix}& (x=0), \\
\\
\dfrac{1}{\sqrt{2}}\begin{bmatrix} 
1 & e^{i\sigma_{-}} \\ 
0 & 0 
\end{bmatrix}& (x\leq-1), \\
\end{array} \right. 
\qquad 
Q_x = \left\{ \begin{array}{ll}
\dfrac{1}{\sqrt{2}}
\begin{bmatrix} 
0 & 0 \\ 
e^{-i\sigma_{+} }& -1
\end{bmatrix}& (x\geq1), \\
\\
\begin{bmatrix} 
0 & 0 \\ 
0 & -1 
\end{bmatrix}& (x=0), \\
\\
\dfrac{1}{\sqrt{2}}
\begin{bmatrix} 
0 & 0 \\ 
e^{-i\sigma_{-} }& -1
\end{bmatrix} & (x\leq-1), \\
\end{array} \right.
\end{align*}
with $U_x = P_x + Q_x$. We should note that $P_x$ and $Q_x$ correspond to the left and right movements, respectively. 
The walker moves differently in positive and negative parts each other. Hereafter, we call the QW ``the two-phase QW''.
Putting $\sigma_{+}=\sigma_{-}$, the model becomes one-defect QW, which has been analyzed so far in detail \cite{segawa}. 
We should note that owing to the defect at the origin, the model has an origin symmetry, and the analysis becomes simple. We will report the analytical results of a QW with two phases which
does not have defect at the origin in the upcoming paper. 
We derived localization theorems \cite{endosan} for the two-phase QW, in particular, the time-averaged limit and stationary measures.
Therefore, by obtaining the weak limit theorem corresponding to the ballistic spreading, we can mathematically express the whole picture of the asymptotic behavior of the two-phase QW with one defect.

\subsection{Weak limit theorem}
Let $C$ be the summation of the time-averaged limit measure $\overline{\mu}_{\infty}(x)$ obtained by Theorem $2$ in \cite{endosan} over all the positions $x\in\mathbb{Z}$, and $X_{t}$ be the quantum walker of the position at time $t$. 
We should note that the time-averaged limit measure describes localization mathematically.
Now, we present 
the weak limit theorem for the missing part $1-C$ with $0\leq C<1$. 
The proof of Theorem \ref{weaklimit} is given in Section \ref{proof_theo}.
In general, the weak limit theorem describes the ballistic spreading of the QW \cite{konnoweak}. 
\par\indent
\begin{theo}
\label{weaklimit}
Let QW be the two-phase model starting from the origin with the initial coin state $\varphi_{0}={}^T\![\alpha,\beta]$, where $\alpha,\beta\in\mathbb{C}$. 
Put $\alpha=ae^{\phi_{1}},\;\beta=be^{\phi_{2}}$ with $a,b\geq0,\;a^{2}+b^{2}=1$ and $\phi_{1}, \phi_{2}\in\mathbb{R}$, where $\mathbb{R}$ is the set of real numbers. Let $\sigma=(\sigma_{+}-\sigma_{-})/2$ and $\tilde{\phi}_{12}=\phi_{1}-\phi_{2}\;$.
For the two-phase QW, $X_{t}/t$ converges weakly to the random variable $Z$ which has the following measure:
\[\mu(dx)=C\delta_{0}(dx)+w(x)f_{K}(x;1/\sqrt{2})dx,\]
where $f_{K}(x;1/\sqrt{2})$ is defined by Eq. \eqref{konnofunction} and
\begin{eqnarray}\label{weight}
w(x)=\dfrac{t_{3}x^{5}+t_{2}x^{4}+t_{1}x^{3}+t_{0}x^{2}}{s_{2}x^{4}+s_{1}x^{2}+s_{0}},
\end{eqnarray}
with
\begin{eqnarray*}s_{2}\!\!\!&=&\!\!\!4\cos^{4}\sigma,\;s_{1}=4\cos^{2}\sigma(1+2\sin^{2}\sigma),\;s_{0}=\cos^{2}2\sigma,\\
t_{3}\!\!\!&=&\!\!\!4\cos^{2}\sigma(b^{2}-a^{2}),\;t_{2}=4[\cos^{2}\sigma(1+\sqrt{2}ab\operatorname{sgn}(x)\cos\gamma(x))+\sqrt{2}ab\operatorname{sgn}(x)\sin\gamma(x)\sin2\sigma],\\t_{1}\!\!\!&=&\!\!\!2(b^{2}-a^{2}),
t_{0}=2\{1+\sqrt{2}ab\operatorname{sgn}(x)\cos\gamma(x)-\sqrt{2}ab\operatorname{sgn}(x)\sin\gamma(x)\sin2\sigma\},
\end{eqnarray*}
and
\begin{eqnarray}\gamma(x)=\left\{ \begin{array}{ll}
\tilde{\phi}_{12}-\sigma_{-}& (x\geq 0), \\
-\tilde{\phi}_{12}+\sigma_{+}& (x<0). \\
\end{array} \right.
\end{eqnarray}
\end{theo}

\noindent
Here we should note that $w(x)f_{K}(x;1/\sqrt{2})$ is an absolutely continuous part. \\
\indent

If $\sigma_{+}=\sigma_{-}$, then, we see from Eq. \eqref{weight} that the weight function is given by
\begin{align}\label{issou}
w(x)=\dfrac{2x^{2}}{1+2x^{2}}\left\{ \begin{array}{ll}
1+\sqrt{2}\Re{(e^{-i\sigma}\alpha\overline{\beta})}+(b^{2}-a^{2})x & (x\geq0), \\
\\
1-\sqrt{2}\Re{(e^{-i\sigma}\alpha\overline{\beta})}+(b^{2}-a^{2})x & (x<0),
\end{array} \right.
\end{align}
which agrees with the result obtained by Theorem $4.1$ in \cite{segawa}. 
Here we should note that the expression of the weight function in Theorem $4.1$ in Ref. \cite{segawa} contains a typo, and the correct transcription is

\[w(x)=\dfrac{|c|^{2}x^{2}}{(|c|^{2}-m)^{2}+(|c|^{2}-m^{2})x^{2}}\left[\gamma(x)-|a_{0}|^{2}\left\{(|\alpha|^{2}-|\beta|^{2})+\dfrac{2\Re{(a_{0}\alpha\overline{b_{0}\beta})}}{|a_{0}|^{2}}\right\}x\right].\]

As we see in Eqs. \eqref{weight} and \eqref{issou}, the two different quantum coins give such complexity to the weight function. 
In our previous paper \cite{endosan}, we reported that the time-averaged distribution of the two-phase QW is symmetric for the origin, however, we emphasize that the weight function $w(x)$, the main result in this paper, is asymmetric, which suggests that the probability distribution has  asymmetry for the origin. One of the interesting future problems is to show the relation in explicit between the topological insulator and the two-phase QW.

\subsection{Example}
\begin{figure}[h]
\begin{minipage}{0.5\hsize}
\centerline{\epsfig{file=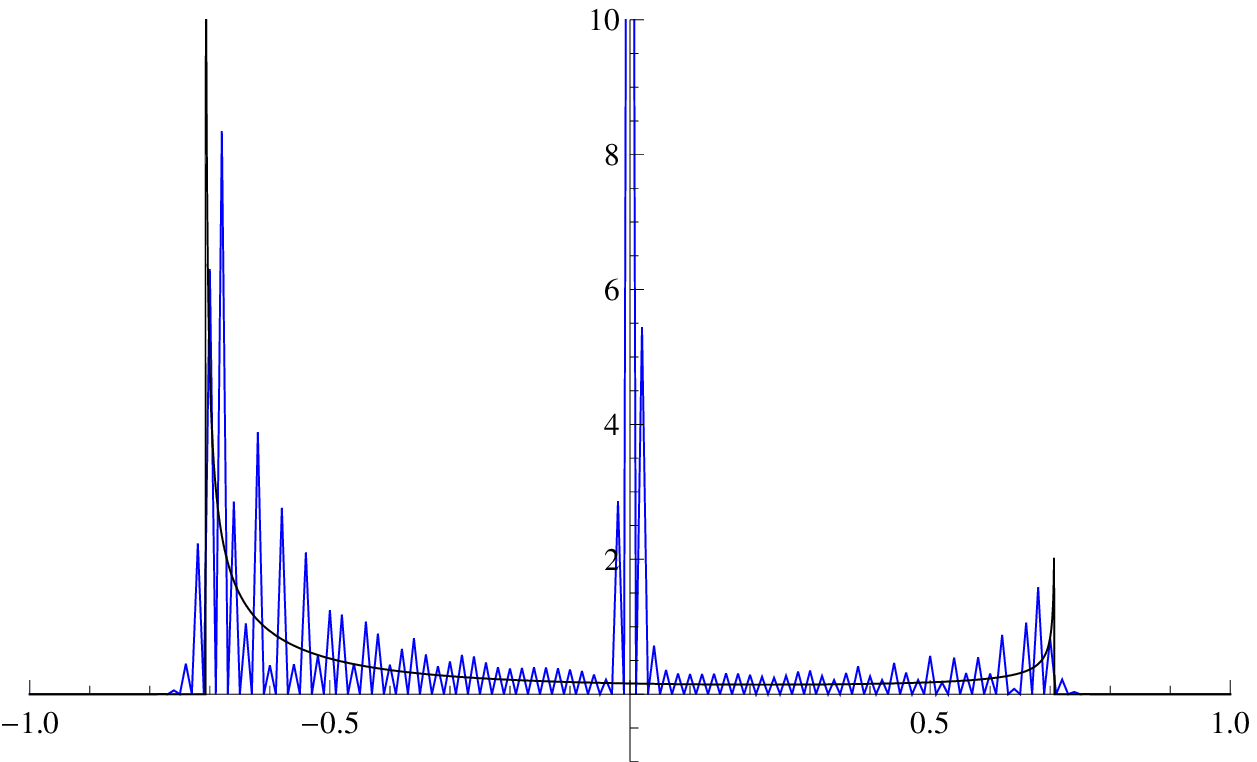, width=7.1cm}} 
\vspace*{13pt}
\fcaption{\label{fig1.} Blue line: Probability distribution in a re-scaled \\ space $(x/100, 100P_{100}(x))$ at time $100$, \\Black line: $w(x)f_{K}(x; 1/\sqrt{2})$}
\end{minipage}
\begin{minipage}{0.5\hsize}
\centerline{\epsfig{file=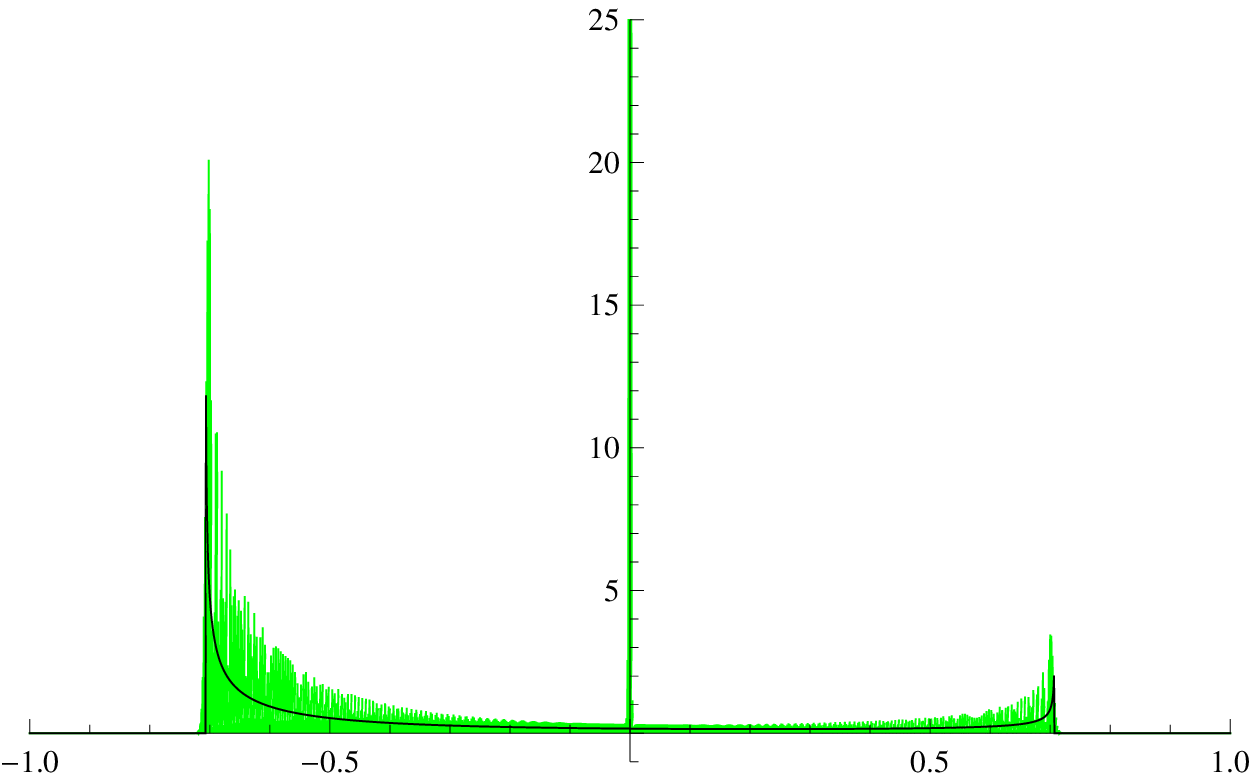, width=7.1cm}} 
\vspace*{13pt}
\fcaption{\label{fig2.}Green line: Probability distribution in a re-scaled \\ space $(x/1000, 1000P_{1000}(x))$ at time $1000$, \\Black line: $w(x)f_{K}(x; 1/\sqrt{2})$}
 \end{minipage}
\end{figure}

\begin{figure}[h]
\begin{minipage}{0.5\hsize}
\centerline{\epsfig{file=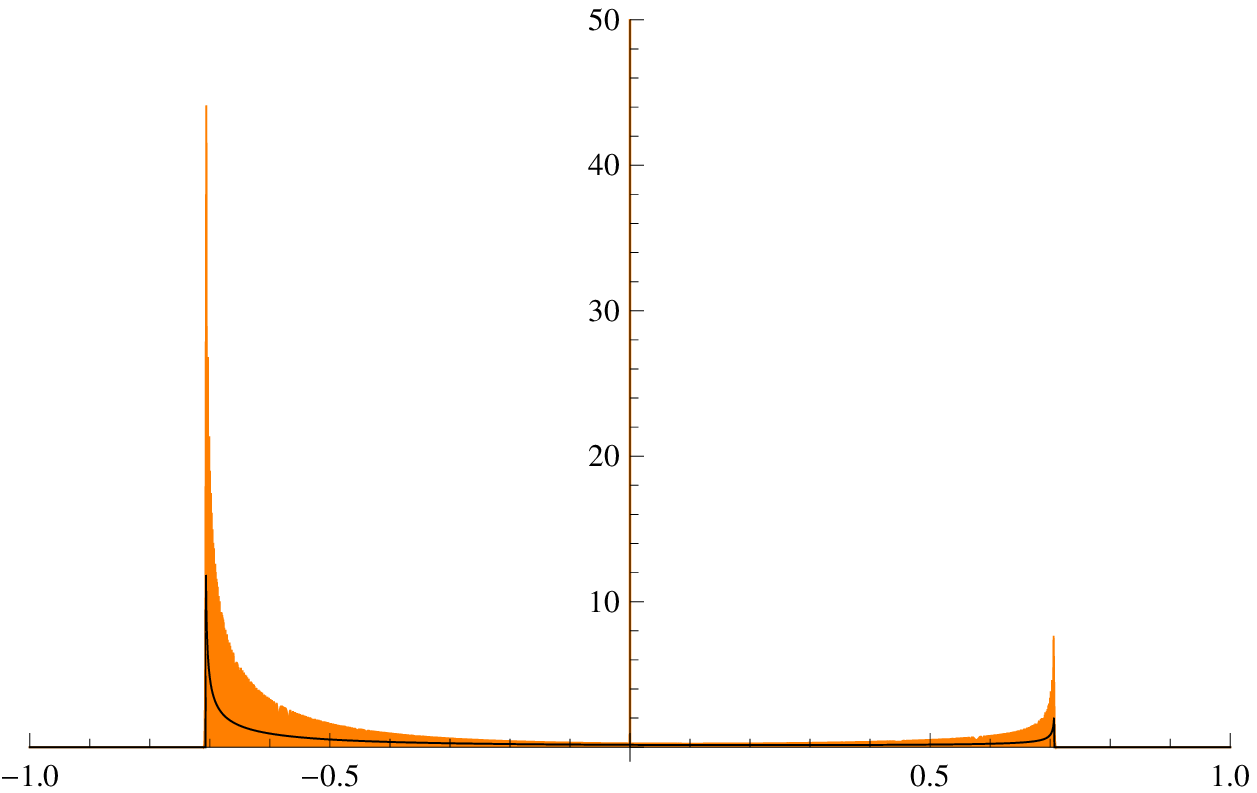, width=7.1cm}} 
\vspace*{13pt}
\fcaption{\label{fig3.}Orange line: Probability distribution in a re-scaled \\ space $(x/10000, 10000P_{1000}(x))$ at time $10000$, \\Black line: $w(x)f_{K}(x; 1/\sqrt{2})$}
 \end{minipage}
\end{figure}

In this subsection, we see a concrete example of our result.
We consider the QW defined by the unitary matrices

\begin{align}
U_{x}=\left\{ \begin{array}{ll}
U_{+}=\dfrac{1}{\sqrt{2}}\begin{bmatrix}1 & -i\\ i & -1 \end{bmatrix}\quad (x=1,2,\cdots), \\
\\
U_{-}=\dfrac{1}{\sqrt{2}}\begin{bmatrix}1 & -1\\ -1 & -1 \end{bmatrix} \quad (x=-1,-2,\cdots),\\
\\
U_{0}=\begin{bmatrix}1 & 0\\ 0 & -1 \end{bmatrix} \quad (x=0).
\end{array} \right.\end{align}
\noindent
We obtain the QW by putting $\sigma_{+}=3\pi/2$ and $\sigma_{-}=\pi$ in Eq. \eqref{2phase_def}.
Let the initial coin state $\varphi_{0}={}^T\![1,0]$.
According to Theorem $1$, the weight function of the QW is
\[w(x)=\dfrac{2(1-x^{3}+x^{2}-x)}{x^{2}+4}.\]
Hence, we see
\begin{align}\int_{-\frac{1}{\sqrt{2}}}^{\frac{1}{\sqrt{2}}}w(x)f_{K}(x;1/\sqrt{2}) dx=\frac{3}{5}.\end{align}
Here, we should note that we obtained the time-averaged limit measure $\overline{\mu}_{\infty}(x)$ by Theorem $2$ in \cite{endosan},  and as a result, we derived the coefficient of the delta function $\delta_{0}(dx)$ in Eq.
\eqref{weakmeasure} by
\begin{align}C=\sum_{x}\overline{\mu}_{\infty}(x)=\frac{4}{25}+2\times\dfrac{12}{25}\sum_{y=1}^{\infty}\left(\frac{1}{5}\right)^{y}=\frac{2}{5},
\end{align}
where 
\begin{align}\overline{\mu}_{\infty}(x)=I_{\{-1/\sqrt{2}\leq\sin\sigma\leq 1\}}(x)\nu^{(+)}(x;\sigma)+I_{\{-1\leq\sin\sigma\leq1/\sqrt{2}\}}(x)\nu^{(-)}(x;\sigma),\label{2-phase.timeaveraged}\end{align}
with
$\tilde{\sigma}=(\sigma_{+}+\sigma_{-})/2,\quad\tilde{\phi}_{12}=\phi_{1}-\phi_{2}$,
and
\begin{eqnarray*}\nu^{(\pm)}(x;\sigma)\!\!\!&=&\!\!\!\left(\dfrac{1\pm\sqrt{2}\sin\sigma}{3\pm2\sqrt{2}\sin\sigma}\right)^{2}\{1\pm2ab\sin(\tilde{\phi}_{12}-\tilde{\sigma})\}\\ 
\!\!\!&\times&\!\!\!\left\{\delta_{0}(x)+(1-\delta_{0}(x))(2\pm\sqrt{2}\sin\sigma)\left(\frac{1}{3\pm2\sqrt{2}\sin\sigma}\right)^{|x|}\right\}.\end{eqnarray*}
Therefore, we have
\[C+\int_{-\frac{1}{\sqrt{2}}}^{\frac{1}{\sqrt{2}}}w(x)f_{K}(x;1/\sqrt{2}) dx=1.\]
Here, we show the numerical results of the probability distribution at time $t=100, 1000$, and $10000$ in re-scaled spaces $(x/t, tP_{t}(x))\;(t=100,1000,10000)$, where $x$ represents the position of the walker and $P_{t}(x)$ is the probability that the walker exists on position $x$ at time $t$. We should note that $x/t$ corresponds to the real axis, and $tP_{t}(x)$ corresponds to the imaginary axis, respectively. Also, we put the graph of  $w(x)f_{K}(x;1/\sqrt{2})$, which is related to absolutely continuous part of the weak limit measure $\mu(dx)$, on the picture at each time. We see that the graph of $w(x)f_{K}(x;1/\sqrt{2})$ is right on the middle of the probability distribution for each position at each time, which suggests that our result is mathematically proper. 
We also emphasize that $\overline{\mu}_{\infty}(x)$ is symmetric for the origin \cite{endosan}, however, $w(x)f_{K}(x;1/\sqrt{2})$ does not have an origin symmetry (Figs. \ref{fig1.}, \ref{fig2.},  \ref{fig3.}) , which indicates that the weak limit measure represents the asymmetry of the probability distribution (Figs. \ref{fig1.}, \ref{fig2.},  \ref{fig3.}).

\section{Proof of Theorem \ref{weaklimit}}
\label{proof_theo}

In this section, we focus on the characteristic function of QW, that is,
\begin{align}E\left[e^{i\xi \frac{X_{t}}{t}}\right]=\int_{x\in\mathbb{Z}}g_{X_{t}/t}(x)e^{i\xi x}dx,\label{density_function}\end{align}
where $g_{X_{t}/t}(x)$ is the density function of random variable $X_{t}/t$. We consider how $E\left[e^{i\xi X_{t}/t}\right]$ can be written when $t\to\infty$.
Here, we should note that to obtain $g_{X_{t}/t}(x)\;(t\to\infty)$ is equivalent to derive $w(x)f_{K}(x;1/\sqrt{2})$.

Let $\Xi_{t}(x)$ be the weight of all the passages of the walker, which moves left $l$ times and moves right $m$ times till time $t\:\cite{segawa}$:
\[\Xi_{t}(x)=\sum_{l_{j},m_{j}}P_{x_{l_{1}}}^{l_{1}}Q_{x_{m_{1}}}^{m_{1}}P_{x_{l_{2}}}^{l_{2}}Q_{x_{m_{2}}}^{m_{2}}\cdots P_{x_{l_{t}}}^{l_{t}}Q_{x_{m_{t}}}^{m_{t}},\]\\
where $l+m=t,\;-l+m=x,\;\;\sum_{i}l_{i}=l,\;\sum_{j}m_{j}=m$ with $l_{i}+m_{i}=1,\;l_{i},m_{i}\in\{0,1\}$, and $\;\sum_{\gamma=l_{i},m_{j}}|x_{\gamma}|=x$. 
Here, we consider $z\in\mathbb{C}$ on a unit circle.
From a simple calculation, we obtain $E[e^{i\xi X_{t}/t}]\;(t\to\infty)$ written by the
square norm of the residue of $\tilde{\Xi}_{x}(z)=\sum_{t}\Xi_{t}(x)z^{t}$ as follows:
\begin{align}
E\left[e^{i\xi\frac{X_{t}}{t}}\right]\to\int^{2\pi}_{0}\sum_{\theta\in A}e^{-i\xi \theta^{'}(k)}\|Res(\hat{\tilde{\Xi}}(k:z):z=e^{i\theta(k)})\|^{2}\frac{dk}{2\pi}\qquad(t\to\infty),\label{limitdensity}
\end{align}
where $A$ is the set of the singular points of $\hat{\tilde{\Xi}}(k:z)\equiv\sum_{x\in\mathbb{Z}}\tilde{\Xi}_{x}(z)e^{ikx}$. Note $\theta^{'}(k)=\partial \theta(k)/\partial k$. 
We will give a detailed explanation of Eq. \eqref{limitdensity} in Appendix A. 
Taking advantage of Eq. \eqref{limitdensity}, we give the proof of Theorem \ref{weaklimit}.

Now, we give useful concrete expressions of $\tilde{\Xi}_{x}(z)$ which play important roles in the proof. 
Lemma \ref{kls} is equivalent to Lemma $2$ in \cite{endosan}, which we used to derive the time-averaged limit measure for the two-phase QW.
Assume that the quantum walker starts from the origin with the initial coin state $\varphi_{0}={}^T\![\alpha,\beta]$ with $\alpha,\beta\in\mathbb{C}$
and $|\alpha|^{2}+|\beta|^{2}=1$.\\
\par\noindent
\begin{lemma}
\label{kls}
$\cite{endosan}$ \label{kls}Let $\Delta_{x}$ be the determinant of $U_{x}$. Assume $a_{x},d_{x}\neq 0$ for all $x\in\mathbb{Z}$.
\begin{enumerate}
\item If $x=0$, we have 
\[\tilde{\Xi}_{0}(z)=\dfrac{1}{1+\tilde{f}_{0}^{(+)}(z)\tilde{f}_{0}^{(-)}(z)}
\begin{bmatrix} 
1 & -\tilde{f}_{0}^{(+)}(z)\\
\tilde{f}_{0}^{(-)}(z) & 1\\
\end{bmatrix}.\]
\item If $|x|\geq 1$, we have
\[
\tilde{\Xi}_{x}(z)=\left\{\begin{array}{ll}
(\tilde{\lambda}^{(+)}(z))^{x-1}
\left[
\begin{array}{c}
\tilde{\lambda}^{(+)}(z)\tilde{f}_{x}^{(+)}(z)\\
z \\
\end{array}
\right][0,-1]\tilde{\Xi}_{0}(z) & (x\geq 1), \\
&\\
(\tilde{\lambda}^{(-)}(z))^{|x|-1}
\left[
\begin{array}{c}
z \\
\tilde{\lambda}^{(-)}(z)\tilde{f}_{x}^{(-)}(z) \\
\end{array}
\right][1,0]\tilde{\Xi}_{0}(z) & (x\leq -1), \\
\end{array} \right.\]
\end{enumerate}
\noindent
where $\tilde{\lambda}^{(+)}(z)=\dfrac{z}{e^{-i\sigma_{+}}\tilde{f}_{0}^{(+)}(z)-\sqrt{2}},\;\tilde{\lambda}^{(-)}(z)=\dfrac{z}{\sqrt{2}-e^{i\sigma_{-}}\tilde{f}_{0}^{(-)}(z)}.$ 
Here $\tilde{f}_{0}^{(+)}(z)$ and $\tilde{f}_{0}^{(-)}(z)$ satisfy the following quadratic equations, respectively:
\[
\left\{
\begin{array}{l}
(\tilde{f}_{x}^{(+)}(z))^{2}-\sqrt{2}e^{i\sigma_{+}}(1+z^{2})\tilde{f}_{x}^{(+)}(z)+e^{2i\sigma_{+}}z^{2}=0,\\
\\
(\tilde{f}_{x}^{(-)}(z))^{2}-\sqrt{2}e^{-i\sigma_{-}}(1+z^{2})\tilde{f}_{x}^{(-)}(z)+e^{-2i\sigma_{-}}z^{2}=0.
\end{array}
\right.
\]
\end{lemma}
Hereafter, we write $\tilde{f}^{(\pm)}_{x}(z)$ by $\tilde{f}_{0}^{(\pm)}(z)$, since $\tilde{f}^{(\pm)}_{x}(z)$ do not depend on the position.
Then, we obtain
\par\indent
\par\noindent
\begin{lemma}
$\tilde{f}_{0}^{(+)}(z)$ and $\tilde{f}_{0}^{(-)}(z)$ are expressed in terms of $\theta$ by
\begin{align}\tilde{f}_{0}^{(\pm)}(e^{i\theta})=e^{i(\theta-\sigma_{\pm})}\times e^{-i\tilde{\phi}(\theta)},\end{align}
where 
\begin{align}\left\{
\begin{array}{l}
\sin\tilde{\phi}(\theta)=\operatorname{sgn}(\sin\theta)\sqrt{2\sin\theta^{2}-1},\\
\cos\tilde{\phi}(\theta)=\sqrt{2}\cos\theta.
\end{array}
\right.\label{tildephi}\end{align} 
\end{lemma}

From now on, we derive the singular points of $\hat{\tilde{\Xi}}(k:z)$ and then, compute the residues of $\hat{\tilde{\Xi}}(k:z)$ at the singular points.
Using Lemma $1$, we can write down $\hat{\tilde{\Xi}}(k:z)$ by
\begin{align}\hat{\tilde{\Xi}}(k:z)=\left\{\frac{e^{ik}}{1-e^{ik}\tilde{\lambda}^{(+)}(z)}\begin{bmatrix}\tilde{\lambda}^{(+)}(z)\tilde{f}_{0}^{(+)}(z)\\ z\end{bmatrix}[0,-1]+\frac{e^{-ik}}{1-e^{-ik}\tilde{\lambda}^{(-)}(z)}\begin{bmatrix}z\\ \tilde{\lambda}^{(-)}(z)\tilde{f}_{0}^{(-)}(z)\end{bmatrix}[1,\;0]+I\right\}\tilde{\Xi}_{0}(z).\label{french}\end{align}
The first term comes from the positive part of $\tilde{\Xi}_{x}(z)$, and the second term comes from the negative part of $\tilde{\Xi}_{x}(z)$, respectively.

Here, we should remark that if $|z|<1$, then $|\tilde{\lambda}^{(\pm)}(z)|<1$. Thus, the infinite series $\sum_{x}(\tilde{\lambda}^{(+)}(z))^{|x|-1}e^{ikx}$ and $\sum_{x}(\tilde{\lambda}^{(-)}(z))^{|x|-1}e^{-ikx}$ converge.
Moreover, as we see in Appendix B, we have 
\begin{align}
\left\{
\begin{array}{l}
\tilde{\lambda}^{(\pm)}(e^{i\theta})=\mp\{\operatorname{sgn}(\cos\theta)\sqrt{2\cos^{2}\theta-1}+i\sqrt{2}\sin\theta\},\\
\\
\tilde{f}_{0}^{(\pm)}(e^{i\theta})=-\operatorname{sgn}(\cos\theta)e^{i(\theta\pm\sigma_{\pm})}\{\sqrt{2}|\cos\theta|-\sqrt{2\cos^{2}\theta-1}\}.
\end{array}
\right.\label{nippon}
\end{align}

We should also note that the singular points derived from $\tilde{\Xi}_{0}(z)$ correspond to localization. On the other hand, the
principal singular points in this paper come from 
\begin{eqnarray}1-e^{ik}\tilde{\lambda}^{(+)}(z)=0,\label{eqt.1}\end{eqnarray}and
\begin{eqnarray}1-e^{-ik}\tilde{\lambda}^{(-)}(z)=0.\label{eqt.2}\end{eqnarray}
\noindent
The solutions of Eqs. \eqref{eqt.1} and \eqref{eqt.2} satisfy the next conditions.
For Eq. \eqref{eqt.1}, we see 
\begin{eqnarray}\cos k=\operatorname{sgn}(\cos\theta^{(+)}(k))\sqrt{2\cos^{2}\theta^{(+)}(k)-1},\label{cosk+}\end{eqnarray}
\begin{eqnarray}\sin k=\sqrt{2}\sin\theta^{(+)}(k),\label{sink+}\end{eqnarray}
and for Eq. \eqref{eqt.2}, we have
\begin{eqnarray}\cos k=-\operatorname{sgn}(\cos\theta^{(-)}(k)(k))\sqrt{2\cos^{2}\theta^{(-)}(k)-1},\label{cosk-}\end{eqnarray}
\begin{eqnarray}\sin k=\sqrt{2}\sin\theta^{(-)}(k).\label{sink-}\end{eqnarray}
\noindent
To compute the RHS of Eq. \eqref{limitdensity} and derive $g_{X_{t}/t}(x)\;(t\to\infty)$ comparing Eq. \eqref{limitdensity} with Eq. \eqref{density_function}, we put $-\partial\theta^{(\pm)}(k)/\partial k=x_{\pm}$.
Then, we derivate Eqs. \eqref{cosk+} and \eqref{cosk-} with respect to $k$, and we obtain $\sin k,\;\cos k,\;\sin\theta^{(\pm)}(k)$, and $\cos\theta^{(\pm)}(k)$ as follows.
From Eqs. \eqref{cosk+} and \eqref{sink+}, we have
\begin{eqnarray}
\left\{
\begin{array}{l}
\cos k=-\operatorname{sgn}(\cos k)\dfrac{x_{+}}{\sqrt{1-x_{+}^{2}}},\;\cos\theta^{(+)}(k)=-\operatorname{sgn}(\cos k)\dfrac{1}{\sqrt{2(1-x_{+}^{2})}},\\
\\
\sin k=\operatorname{sgn}(\sin k)\sqrt{\dfrac{1-2x_{+}^{2}}{1-x_{+}^{2}}},\;\;\sin\theta^{(+)}(k)=\operatorname{sgn}(\sin k)\sqrt{\dfrac{1-2x_{+}^{2}}{2(1-x_{+}^{2})}}.
\end{array}
\right.\label{solutions+}
\end{eqnarray} 
From Eqs. \eqref{cosk-} and \eqref{sink-}, we see
\begin{eqnarray}
\left\{
\begin{array}{l}
\cos k=\operatorname{sgn}(\cos k)\dfrac{x_{-}}{\sqrt{1-x_{-}^{2}}},\;\cos\theta^{(-)}(k)=\operatorname{sgn}(\cos k)\dfrac{1}{\sqrt{2(1-x_{-}^{2})}},\\
\sin k=\operatorname{sgn}(\sin k)\sqrt{\dfrac{1-2x_{-}^{2}}{1-x_{-}^{2}}},\;\;\sin\theta^{(-)}(k)=\operatorname{sgn}(\sin k)\sqrt{\dfrac{1-2x_{-}^{2}}{2(1-x_{-}^{2})}}.
\end{array}
\right.\label{solutions-}
\end{eqnarray} 
Therefore, we obtain the set of the singular points of $\hat{\tilde{\Xi}}(k:z)$, A, as follows:
\[A=\{e^{i\theta^{(+)}(k)},e^{i\theta^{(-)}(k)}\},\]
where
\[e^{i\theta^{(+)}(k)}=\frac{\operatorname{sgn}(\cos k)}{\sqrt{2(1-x^{2}_{+})}}+i\operatorname{sgn}(\sin k)\sqrt{\frac{1-2x^{2}_{+}}{2(1-x^{2}_{+})}},\]
and 
\[e^{i\theta^{(-)}(k)}=-\frac{\operatorname{sgn}(\cos k)}{\sqrt{2(1-x^{2}_{-})}}+i\operatorname{sgn}(\sin k)\sqrt{\frac{1-2x^{2}_{-}}{2(1-x^{2}_{-})}}.\]

In the next stage, we derive the residue of $\hat{\tilde{\Xi}}(k;z)$ at $e^{i\theta^{(\pm)}(k)}$.
At first, substituting the singular points to $\tilde{f}_{0}^{(\pm)}(z)$, we obtain\\
\begin{enumerate}
\item $\tilde{f}_{0}^{(+)}(e^{i\theta^{(+)}(k)})=-\operatorname{sgn}(\cos k)e^{i(\theta^{+}(k)+\sigma_{+})}\dfrac{\sqrt{1-x^{2}}}{1+|x|},\;\;\tilde{f}_{0}^{(-)}(e^{i\theta^{(+)}(k)})=-\operatorname{sgn}(\cos k)e^{i(\theta^{(+)}(k)-\sigma_{-})}\dfrac{\sqrt{1-|x|^{2}}}{1+|x|}$,\\
\item $\tilde{f}_{0}^{(+)}(e^{i\theta^{(-)}(k)})=\operatorname{sgn}(\cos k)e^{i(\theta^{(-)}(k)+\sigma_{+})}\dfrac{\sqrt{1-x^{2}}}{1+|x|},\;\;\tilde{f}_{0}^{(-)}(e^{i\theta^{(-)}(k)})=\operatorname{sgn}(\cos k)e^{i(\theta^{(-)}(k)-\sigma_{-})}\dfrac{\sqrt{1-|x|^{2}}}{1+|x|}$.\\
\end{enumerate}
\noindent
\par\noindent
Noting Lemma \ref{kls}, we see
\[\frac{e^{ik}}{1-e^{ik}\tilde{\lambda}^{(+)}(z)}\begin{bmatrix}\tilde{f}_{0}^{(+)}(z)\tilde{\lambda}^{(+)}(z)\\ z\end{bmatrix}[0,\;-1]\tilde{\Xi}_{0}(z)=\frac{1}{\tilde{\Lambda}_{0}(z)}\frac{e^{ik}}{1-e^{ik}\tilde{\lambda}^{(+)}(z)}\begin{bmatrix}\tilde{f}_{0}^{(+)}(z)\tilde{\lambda}^{(+)}(z)\\ -z\end{bmatrix}(\alpha\tilde{f}_{0}^{(-)}(z)+\beta),\]
\par\noindent
and the square norm of residue of the first term of Eq. \eqref{french} is written as\\

$\left|Res\left(\frac{e^{ik}}{1-e^{ik}\tilde{\lambda}^{(+)}(z)}\begin{bmatrix}\tilde{f}_{0}^{(+)}(z)\tilde{\lambda}^{(+)}(z)\\ z\end{bmatrix}[0,\;-1]\tilde{\Xi}_{0}(z):z=e^{i\theta^{(+)}(k)}\right)\right|^{2}$
\begin{eqnarray*}&=&\!\!\!\!\left|Res\left(\frac{1}{1-e^{ik}\tilde{\lambda}^{(+)}(z)}:z=e^{i\theta^{(+)}(k)}\right)\right|^{2}\frac{1}{|\tilde{\Lambda}_{0}(e^{i\theta^{(+)}(k)})|^{2}}\left|\begin{bmatrix}\tilde{f}_{0}^{(+)}(e^{i\theta^{(+)}(k)})\tilde{\lambda}^{(+)}(e^{i\theta^{(+)}(k)})\\ -e^{i\theta^{(+)}(k)}\end{bmatrix} \right|^{2}\\
&\times&\!\!\!\!|\alpha\tilde{f}_{0}^{(-)}(e^{i\theta^{(+)}(k)})+\beta|^{2}.
\end{eqnarray*}
\par\indent
\par\noindent
In a similar way, we can write down the second term of Eq. \eqref{french} by

$\left|Res\left(\frac{e^{-ik}}{1-e^{-ik}\tilde{\lambda}^{(-)}(z)}\begin{bmatrix}z\\ \tilde{f}_{0}^{(-)}(z)\tilde{\lambda}^{(-)}(z)\end{bmatrix}[1,\;0]\tilde{\Xi}_{0}(z):z=e^{i\theta^{(-)}(k)}\right)\right|^{2}$
\begin{eqnarray*}&=&\!\!\!\!\left|Res\left(\frac{1}{1-e^{-ik}\tilde{\lambda}^{(-)}(z)}:z=e^{i\theta^{(-)}(k)}\right)\right|^{2}\frac{1}{|\tilde{\Lambda}_{0}(e^{i\theta^{(-)}(k)})|^{2}}\left|\begin{bmatrix}e^{i\theta^{(-)}(k)}\\ \tilde{f}_{0}^{(-)}(e^{i\theta^{(-)}(k)}) \tilde{\lambda}^{(-)}(e^{i\theta^{(-)}(k)})\end{bmatrix}
\right|^{2}\\&\times&\!\!\!\!|\alpha-\beta\tilde{f}_{0}^{(+)}(e^{i\theta^{(-)}(k)})|^{2}.\end{eqnarray*}

\par\indent
\par\noindent
Hence, we obtain
\begin{eqnarray*}
\|Res(\hat{\tilde{\Xi}}(k:z)\!\!\!\!\!&:&\!\!\!\!\!z=e^{i\theta^{(\pm)}(k)})\|^{2}
=\left|Res\left(\dfrac{1}{1-e^{ik}\tilde{\lambda}^{(+)}(z)}:z=e^{i\theta^{(+)}(k)}\right)\right|^{2}\dfrac{1}{|\tilde{\Lambda}_{0}(e^{i\theta^{(+)}(k)})|^{2}}\\ 
\!\!\!\!\!&\times&\!\!\!\!\!\!\left|\begin{bmatrix}\tilde{f}_{0}^{(+)}(e^{i\theta^{(+)}(k)})\tilde{\lambda}^{(+)}(e^{i\theta^{(+)}(k)}) \\-e^{i\theta^{(+)}(k)}\end{bmatrix}\right|^{2}|\alpha\tilde{f}_{0}^{(-)}(e^{i\theta^{(+)}(k)})+\beta|^{2}\\
\!\!\!\!&+&\!\!\!\left|Res\left(\dfrac{1}{1-e^{-ik}\tilde{\lambda}^{(-)}(z)}:z=e^{i\theta^{(-)}(k)}\right)\right|^{2}\frac{1}{|\tilde{\Lambda}_{0}(e^{i\theta^{(-)}(k)})|^{2}}\\
\!\!\!&\times&\!\!\!\left|\begin{bmatrix}e^{(i\theta^{(\pm)}(k))}\\ \tilde{f}_{0}^{(-)}(e^{i\theta^{(-)}(k)})\tilde{\lambda}^{(-)}(e^{i\theta^{(-)}(k)})\end{bmatrix}\right|^{2}|\alpha-\beta\tilde{f}_{0}^{(+)}(e^{i\theta^{(-)}(k)})|^{2}.\label{kekka}
\end{eqnarray*}
\par\indent
\par\noindent
Henceforth, we will express the items below in terms of $x_{+}$ or $x_{-}$, and then substitute the items in Eq. \eqref{kekka}.

\begin{enumerate}
\item $\left|Res\left(\dfrac{1}{1-e^{ik}\tilde{\lambda}^{(+)}(z)}: z=e^{i\theta^{(+)}(k)}\right)\right|^{2}$ and $\left|Res\left(\dfrac{1}{1-e^{-ik}\tilde{\lambda}^{(-)}(z)}: z=e^{i\theta^{(-)}(k)}\right)\right|^{2}$,\\
\item $\dfrac{1}{|\tilde{\Lambda}_{0}(e^{i\theta^{(\pm)}(k)})|^{2}}$,\\
\item $|\alpha\tilde{f}_{0}^{(-)}(e^{i\theta^{(+)}(k)})+\beta|^{2}$ and $|\alpha-\beta\tilde{f}_{0}^{(+)}(e^{i\theta^{(-)}(k)})|^{2}$,\\
\item $\left\|\begin{bmatrix}\tilde{\lambda}^{(+)}(e^{i\theta^{(+)}(k)})\tilde{f}_{0}^{(+)}(e^{i\theta^{(+)}(k)})\\ -e^{i\theta^{(+)}(k)}\end{bmatrix}\right\|^{2}$ and $\left\|\begin{bmatrix}e^{i\theta^{(-)}(k)}\\ \tilde{\lambda}^{(-)}(e^{i\theta^{(-)}(k)})\tilde{f}_{0}^{(-)}(e^{i\theta^{(-)}(k)})\end{bmatrix}\right\|^{2}$.\\
\end{enumerate}
\par\noindent
$1.$ Computation of $\left|Res\left(\dfrac{1}{1-e^{ik}\tilde{\lambda}^{(+)}(z)}: z=e^{i\theta^{(+)}(k)}\right)\right|^{2}$ and $\left|Res\left(\dfrac{1}{1-e^{-ik}\tilde{\lambda}^{(-)}(z)}: z=e^{i\theta^{(-)}(k)}\right)\right|^{2}$.\\
\par\noindent
Let $g^{(\pm)}(z)=1-e^{\pm ik}\tilde{\lambda}^{(\pm)}(z)$. Expanding $g^{(\pm)}(z)$ around $z=e^{i\theta^{(\pm)}(k)}$, we have
\[Res\left(\frac{1}{1-e^{\pm ik}\tilde{\lambda}^{(\pm)}(z)}:z=e^{i\theta^{(\pm)}(k)}\right)=\left.\frac{1}{\dfrac{\partial g^{(\pm)}(z)}{\partial z}}\right|_{z=e^{i\theta^{(\pm)}(k)}}.\]
From Eqs. \eqref{nippon}, we see
\[\left.\frac{\partial g^{(\pm)}(z)}{\partial z}\right|_{z=e^{i\theta^{(\pm)}(k)}}=\pm\dfrac{\operatorname{sgn}(\cos k)}{\sqrt{1-x^{2}}}e^{-i(\theta^{(\pm)}(k)\mp k)}\left\{\operatorname{sgn}(\cos k\sin k)\sqrt{1-2x^{2}}+i\right\},\]
which imply
\begin{eqnarray}
\left\{
\begin{array}{l}
\left|Res\left(\dfrac{1}{1-e^{ik}\tilde{\lambda}^{(+)}(z)} : z=e^{i\theta^{(\pm)}(k)}\right)\right|^{2}=x_{+}^{2},\\
\\
\left|Res\left(\dfrac{1}{1-e^{-ik}\tilde{\lambda}^{(-)}(z)}: z=e^{i\theta^{(\pm)}(k)}\right)\right|^{2}=x_{-}^{2}.
\end{array}
\right.
\end{eqnarray} 
\noindent
$2.$ Computation of $\dfrac{1}{|\tilde{\Lambda}_{0}(e^{i\theta^{(\pm)}(k)})|^{2}}$.\\
\par\noindent
Noting Lemma \ref{kls}, we have for any $\theta\in\mathbb{R}$, 
\begin{align}
|\tilde{\Lambda}_{0}(e^{i\theta})|^{2}=1+2Re\{\tilde{f}_{0}^{(+)}(e^{i\theta})\tilde{f}_{0}^{(-)}(e^{i\theta})\}+|\tilde{f}_{0}^{(+)}(e^{i\theta})|^{2}|\tilde{f}_{0}^{(-)}(e^{i\theta})|^{2},\label{tokyo}
\end{align}
where $\mathbb{R}$ is the set of the real numbers.
Hence, substituting the singular points into Eq. \eqref{tokyo}, we obtain
\begin{eqnarray}
\left\{
\begin{array}{l}
\left|\dfrac{1}{\tilde{\Lambda}_{0}(e^{i\theta^{(+)}(k)})}\right|^{2}=\dfrac{(1+x_{+})^{2}}{2\{1+x_{+}^{2}(1+\cos 2\sigma)+\operatorname{sgn}(\sin k\cos k)\sqrt{1-2x_{+}^{2}}\sin 2\sigma\}},\\
\\
\left|\dfrac{1}{\tilde{\Lambda}_{0}(e^{i\theta^{(-)}(k)})}\right|^{2}=\dfrac{(1-x_{-})^{2}}{2\{1+x_{-}^{2}(1+\cos 2\sigma)-\operatorname{sgn}(\sin k\cos k)\sqrt{1-2x_{-}^{2}}\sin 2\sigma\}}.
\end{array}
\right.
\end{eqnarray} 
\noindent
$3.$ Computation of $|\alpha\tilde{f}_{0}^{(-)}(e^{i\theta^{(+)}(k)})+\beta|^{2}$ and $|\alpha-\beta\tilde{f}_{0}^{(+)}(e^{i\theta^{(-)}(k)})|^{2}$.\\
\par\noindent
Let the initial coin state $\varphi_{0}={}^T\![\alpha,\beta]$, where $\alpha=ae^{i\phi_{1}},\;\beta=be^{i\phi_{2}}$ with $a,b\geq0$ and $a^{2}+b^{2}=1$. 
Noting 
\[|\alpha\tilde{f}_{0}^{(-)}(e^{i\theta^{(+)}(k)})+\beta|^{2}=|\alpha|^{2}|\tilde{f}_{0}^{(-)}(e^{i\theta^{(+)}(k)})|^{2}+|\beta|^{2}+2\Re\{\alpha\overline{\beta}\tilde{f}_{0}^{(-)}(e^{i\theta^{(+)}(k)})\},\]
and
\[|\alpha-\beta\tilde{f}_{0}^{(+)}(e^{i\theta^{(-)}(k)})|^{2}=|\alpha|^{2}-2\Re\{\overline{\alpha}\beta\tilde{f}_{0}^{(-)}(e^{i\theta^{(-)}(k)})\}+|\beta|^{2}|\tilde{f}_{0}^{(+)}(e^{i\theta^{(-)}(k)})|^{2},\]
we obtain
\begin{eqnarray}
\left\{
\begin{array}{l}
|\alpha\tilde{f}_{0}^{(-)}(e^{i\theta^{(+)}(k)})+\beta|^{2}=a^{2}\dfrac{(1-x_{+})}{1+x_{+}}+b^{2}+\dfrac{\sqrt{2}ab}{1-x_{+}}\{\cos\gamma_{+}+\operatorname{sgn}(\sin k\cos k)\sqrt{1-2x_{+}^{2}}\sin\gamma_{+}\},\\
|\alpha-\beta\tilde{f}_{0}^{(+)}(e^{i\theta^{(-)}(k)})|^{2}=a^{2}-\dfrac{\sqrt{2}ab}{1-x_{-}}\{\cos\gamma_{-}-\operatorname{sgn}(\sin k\cos k)\sqrt{1-2x_{-}^{2}\sin\gamma_{-}}\}+b^{2}\dfrac{1+x_{-}}{1-x_{-}},
\end{array}
\right.
\end{eqnarray} 
where
$\gamma_{+}=\tilde{\phi}_{12}-\sigma_{-}$ and $\gamma_{-}=\tilde{\phi}_{21}+\sigma_{+}$ with $\tilde{\phi}_{12}=\phi_{1}-\phi_{2}$.\\
\par\noindent
$4.$ Computation of $\left\|\begin{bmatrix}\tilde{\lambda}^{(+)}(e^{i\theta^{(+)}(k)})\tilde{f}_{0}^{(+)}(e^{i\theta^{(+)}(k)})\\ -e^{i\theta^{(+)}(k)}\end{bmatrix}\right\|^{2}$ and $\left\|\begin{bmatrix}e^{i\theta^{(-)}(k)}\\ \tilde{\lambda}^{(-)}(e^{i\theta^{(-)}(k)})\tilde{f}_{0}^{(-)}(e^{i\theta^{(-)}(k)})\end{bmatrix}\right\|^{2}$.\\
\par\noindent
By a simple calculation, we have
\begin{eqnarray}
\left\{
\begin{array}{ll}
\left\|\begin{bmatrix}\tilde{\lambda}^{(+)}(e^{i\theta^{(+)}(k)})\tilde{f}_{0}^{(+)}(e^{i\theta^{(+)}(k)})\\ -e^{i\theta^{(+)}(k)}\end{bmatrix}\right\|^{2}=|\tilde{\lambda}^{(+)}(e^{i\theta^{(+)}(k)})|^{2}|\tilde{f}_{0}^{(+)}(e^{i\theta^{(+)}(k)})|^{2}+1=\dfrac{2}{1+x_{+}}&(x_{+}>0),\\
\left\|\begin{bmatrix}e^{i\theta^{(-)}(k)}\\ \tilde{\lambda}^{(-)}(e^{i\theta^{(-)}(k)})\tilde{f}_{0}^{(-)}(e^{i\theta^{(-)}(k)})\end{bmatrix}\right\|^{2}=1+|\tilde{\lambda}^{(-)}(e^{i\theta^{(-)}(k)})|^{2}|\tilde{f}_{0}^{(-)}(e^{i\theta^{(-)}(k)})|^{2}=\dfrac{2}{1-x_{-}}&(x_{-}<0).
\end{array}
\right.
\end{eqnarray}
\\
\par\noindent
Here, we should remark 
\begin{align}-\frac{\partial\theta^{(\pm)}(k)}{\partial k}=x_{\pm}.\label{hensu}\end{align}
Eq. \eqref{hensu} implies
\begin{align}x_{+}=\frac{|\cos k|}{\sqrt{1+\cos^{2}k}},\;\;\;
x_{-}=-\frac{|\cos k|}{\sqrt{1+\cos^{2}k}}.\label{x}\end{align}
Hence, we can regard $x_{+}$ and $x_{-}$ as a variable $x$; 
\[
x=\left\{ \begin{array}{ll}
x_{+} & (x>0), \\
x_{-} & (x<0).
\end{array} \right.\] 
\par\noindent
Combining Eqs. \eqref{solutions+} and \eqref{solutions-} with Eq. \eqref{x}, and noting Eq. (\eqref{hensu}), we get
\[\frac{dx}{dk}=\operatorname{sgn}(x)\operatorname{sgn}(\sin k\cos k)(1-x^{2})\sqrt{1-2x^{2}},\]
and therefore, we obtain
\begin{align}
dk=\left\{ \begin{array}{ll}
\operatorname{sgn}(\sin k\cos k)f_{K}(x;1/\sqrt{2})\pi dx&(x>0), \\
-\operatorname{sgn}(\sin k\cos k)f_{K}(x;1/\sqrt{2})\pi dx& (x<0).
\end{array} \right.
\end{align} 
\noindent
Substituting the items given in $1.$ to $4.$ into Eq. \eqref{kekka} and combining with Eq. \eqref{limitdensity}, we arrive at Theorem \ref{weaklimit}.


\setcounter{footnote}{0}
\renewcommand{\thefootnote}{\alph{footnote}}
\nonumsection{Acknowledgments} 
SE acknowledges financial support of Postdoctoral Fellowship for Research Abroad from Japan Society for the Promotion of Science.
NK acknowledges financial support of the Grant-in-Aid for Scientific
Research (C) of Japan Society for the Promotion of Science (No.21540116).
ES's work is also partially supported by the Grant-in-Aid for young Scientists (B) of Japan Society for the promotion of Science (No.25800088).\\
\par\noindent
\nonumsection{References}

\vspace{-08mm}
\flushleft
\begin{appendix}\\ \noindent
In Appendix A, we explain how Eq. \eqref{limitdensity}, which is a key relation of the proof of Theorem \ref{weaklimit}, is derived. Put 
$w_{l}(k)=Res(\hat{\tilde{\Psi}}_{t}(k:z): z=e^{i\theta_{l}(k)})$ with $\Psi_{t}(x)=\Xi_{t}(x)\varphi_{0}$.
By definition, we have
\begin{eqnarray}E\left[e^{i\xi \frac{X_{t}}{t}}\right]\!\!&=&\!\!\sum_{j}P(X_{t}=j)e^{i\xi\frac{j}{t}}\nonumber\\
\!\!\!&=&\!\!\!\sum_{j}\|\Xi_{t}(j)\varphi_{0}\|^{2}e^{i\xi\frac{j}{t}}\nonumber\\
\!\!\!&=&\!\!\!\sum_{x,y}\varphi_{0}^{\ast}\Xi_{t}^{\ast}(y)\Xi_{t}(x)\varphi_{0}e^{i\xi\frac{x}{t}}\int^{2\pi}_{0}e^{i(x-y)k}\dfrac{dk}{2\pi} \nonumber\\
\!\!&=&\!\!\sum_{x,y}\left<\Psi_{t}(y), \Psi_{t}(x)\right>e^{i\xi\frac{x}{t}}\int^{2\pi}_{0}e^{ik(x-y)}\dfrac{dk}{2\pi}\nonumber\\
\!\!\!&=&\!\!\!\int^{2\pi}_{0}\left<\hat{\Psi}_{t}(k), \hat{\Psi}_{t}\left(k+\dfrac{\xi}{t}\right)\right>\dfrac{dk}{2\pi}\label{banana}\\
\!\!\!&=&\!\!\!\int^{2\pi}_{0}\left<\sum_{l}w_{l}(k)e^{-i(t+1)\theta_{l}(k)},\sum_{m}w_{m}\left(k+\dfrac{\xi}{t}\right)e^{-i(t+1)\theta_{m}(k+\frac{\xi}{t})}\right>\dfrac{dk}{2\pi}\label{ringo}\\
\!\!\!&=&\!\!\!\int^{2\pi}_{0}\left\{\sum_{l}|w_{l}(k)|^{2}e^{-i\xi\frac{t+1}{t}\theta^{`}(k)} e^{-i(t+1)O(\frac{1}
{t^{2}})}+O\left(\frac{1}{t}\right)\right\}\dfrac{dk}{2\pi}\nonumber\\
\!\!\!&+&\!\!\!\int^{2\pi}_{0} \left\{\sum_{l}\sum_{m}\!\!\overline{w_{l}(k)}e^{i(t+1)\theta_{l}(k)}w_{m}(k)e^{-i(t+1)\theta_{m}(k)}e^{-i\xi\frac{t+1}{t}\theta^{`}(k)} e^{-i(t+1) O(\frac{1}{t^{2}})}+O\left(\frac{1}{t}\right)\right\}\dfrac{dk}{2\pi}.\nonumber\\
\label{syomei}\end{eqnarray}
Here we should note that we use the residue theorem when we calculate Eq. \eqref{banana} to Eq. \eqref{ringo}, and Maclaurin's expansion for $w_{m}(k+\xi/t)$ when we calculate Eq. \eqref{ringo} to Eq. \eqref{syomei}.
By the Riemann-Lebesgue Theorem, the second term of Eq. \eqref{syomei} vanishes when $t\to\infty$, and we get the desired equation.
\end{appendix}
\flushleft
\begin{appendix}\\ \noindent
In Appendix B, we consider how $\tilde{f}_{x}^{(\pm)}(z)$ and $\tilde{\lambda}^{(\pm)}$ are fixed when we focus on the ballistic behavior of the two-phase QW. 
According to \cite{endosan}, we have
\[
\left\{
\begin{array}{l}
\tilde{\lambda}^{(\pm)}(w)=\pm\dfrac{i}{\sqrt{2}}\{(w+w^{-1})-\sqrt{(w+w^{-1})^{2}-2}\},\\
\tilde{f}_{0}^{(\pm)}(w)=-\frac{w e^{i\sigma_{+}}}{\sqrt{2}}\{(w-w^{-1})+\sqrt{(w-w^{-1})^{2}+2}\}.
\end{array}
\right.\]

Putting $w=i(1-\epsilon)e^{i\theta}\;(\epsilon\in\mathbb{R},\;|\epsilon|\ll1)$, we consider how $\lim_{\epsilon\to0}\sqrt{(w+w^{-1})^{2}-2}$ can be specified in terms of $\theta$ according to the range of $\cos\theta$ or $\sin\theta$.
Noting $|\epsilon|\ll1$, we can approximates $\tilde{\lambda}^{(\pm)}(w)$ as \cite{endosan} 
\begin{eqnarray}
\tilde{\lambda}^{(+)}(w)\!\!\!&=&\!\!\!\frac{i}{\sqrt{2}}\left\{(1-\epsilon)ie^{i\theta}-(1-\epsilon)^{-1}ie^{-i\theta}-\sqrt{\{(1-\epsilon)ie^{i\theta}-(1-\epsilon)^{-1}ie^{-i\theta}\}^{2}-2}\right\}\nonumber\\
\!\!\!&\sim&\!\!\!-\frac{i}{\sqrt{2}}\left\{2\sin\theta+2i\epsilon\cos\theta+\delta\sqrt{4\sin^{2}\theta-2}\right\},\label{gyoten}\end{eqnarray}
where we put $\delta\in\mathbb{R}$ and $\delta^{2}=1$. 
Noting $|\tilde{\lambda}^{(+)}(w)|<1$,
Eq. \eqref{gyoten} suggests that we need to take into consideration the next two cases \cite{endosan}.
\begin{enumerate}
\item$|\sin\theta|\geq 1/\sqrt{2}$ case.\\
Eq. \eqref{gyoten} implies  
\[\frac{1}{2}\left\{2\sin\theta+2\delta\sqrt{\sin^{2}\theta-1/2}\right\}^{2}<1.\]
Thus, we have
\[2\sin^{2}\theta+2\sin\theta\delta\sqrt{\sin^{2}\theta-1/2}<1.\]
Consequently, we obtain $\delta=-\operatorname{sgn}(\sin\theta)$.
\item$|\sin\theta|<1/\sqrt{2}$ case.\\
Eq. \eqref{gyoten} also implies 
\[\frac{1}{2}\left[\left\{2\sin\theta+2\delta\sqrt{\sin^{2}\theta-1/2}\right\}^{2}+4\epsilon^{2}\cos^{2}\theta\right]<1.\]
Thus, we see
\[4\epsilon^{2}\cos^{2}\theta+8\epsilon\cos\theta\delta\sqrt{1/2-\sin^{2}\theta}<0.\]
Consequently, we obtain $\delta=-\operatorname{sgn}(\cos\theta)$.
\end{enumerate}
As a result, the square root is expressed as
\begin{align}\lim_{\epsilon\to0}\sqrt{(w+w^{-1})^{2}-2}=\left\{
\begin{array}{ll}
-2\operatorname{sgn}(\sin\theta)\sqrt{\sin^{2}\theta-\frac{1}{2}}&(\;|\sin\theta|\geq1/\sqrt{2}\;),\\
-2i\operatorname{sgn}(\cos\theta)\sqrt{\frac{1}{2}-\sin^{2}\theta}&(\;|\sin\theta|\leq1/\sqrt{2}\;).
\end{array}
\right.\label{news}\end{align}
Next, we determine in detail $\tilde{\lambda}^{(\pm)}(z)$ and $\tilde{f}_{0}^{(\pm)}(z)$.
When we focus on the weak limit theorem for our two-phase QW, we choose the square root so that $1/(1-e^{ik}\tilde{\lambda}^{(+)}(z))$ and $1/(1-e^{-ik}\tilde{\lambda}^{(-)}(z))$ have the singular points, that is, $|\tilde{f}_{0}^{(\pm)}(z)|\neq1$.
Hence  Eq. \eqref{news} gives
\[\left\{
\begin{array}{l}
\tilde{\lambda}^{(\pm)}(z)=\mp\{\operatorname{sgn}(\cos\theta)\sqrt{2\cos^{2}\theta-1}+i\sqrt{2}\sin\theta\},\\
\\
\tilde{f}_{0}^{(\pm)}(z)=\operatorname{sgn}(\cos\theta)e^{i(\theta\pm\sigma_{\pm})}\{\sqrt{2}|\cos\theta|-\sqrt{2\cos^{2}\theta-1}\},
\end{array}
\right.\;(|\sin\theta|\leq1/\sqrt{2})
\]
where $z=e^{i\theta}$.\\
\end{appendix}


\vspace{-7mm}
\begin{thebibliography}{99}

\bibitem{scholz}A. Ahlbrecht, V. B. Scholz, and A. H. Werner: Disordered quantum walks in one lattice dimension, Journal of Mathematical Physics, {\bf 52}, 102201 (2011)

\bibitem{ambainis}A. Ambainis, A. Backurs, N. Nahimovs, R. Ozols, and A. Rivosh: Search by quantum walks on two-dimensional grid without amplitude amplification, Springer, Lecture Notes in Computer Science, {\bf 7582}, 87-97 (2013)

\bibitem{be}A. A. Burkov and L. Balents: Weyl Semimetal in a Topological insulator multilayer, Physical Review Letters, {\bf 107}, 127205 (2011)

\bibitem{cantero}M. J. Cantero, F. A. Grunbaum, L. Moral, and L. Velazquez: One-dimensional quantum walks with one defect, Reviews in Mathematical Physics, {\bf 24}, 1250002 (2012)


\bibitem{chen}Y. L. Chen, J. G. Analytis, J.-H. Chu, Z. K. Liu, S.-K. Mo, X. L. Qi, H. J. Zhang, D. H. Lu, X. Dai, Z. Fang, S. C. Zhang, I. R. Fisher, Z. Hussain and Z.-X. Shen: Experimental realization of a three-dimensional topological insulator, Science, {\bf 325}, 178-181 (2009) 

\bibitem{kota}N. Konno: Quantum walks, Quantum Potential Theory,
Lecture Notes in Mathematics, {\bf1954}, 309-452. Springer, Heidelberg (2008)

\bibitem{cho}C.-I. Chou and C.-L. Ho: Localization and recurrence of quantum walk in periodic potential on a line, arXiv:1307.3186 

\bibitem{watanabe}T. Endo and N. Konno: The stationary measure of a space-inhomogeneous quantum walk on the line, Yokohama Mathematical Journal, {\bf 60}, 33-47 (2014)

\bibitem{endo}T. Endo and N. Konno: The time-averaged limit measure of the Wojcik model, Quantum Information and Computation, {\bf 15}, 0105-0133 (2015)

\bibitem{endosan}S. Endo, T. Endo, N. Konno, E. Segawa, and M. Takei: Limit theorems of a two-phase quantum walk with one-defect, arXiv:1409.8134 

\bibitem{fu}L. Fu and C. L. Kane: Superconducting Proximity Effect and Majorana fermions at the surface of a topological insulator, Physical Review Letters, {\bf 100}, 096407 (2008)

\bibitem{inui}N. Inui, Y. Konishi, and N. Konno: Localization of two-dimensional quantum walks,
Physical Review A, {\bf 69}, 052323 (2003)

\bibitem{joye}A. Joye and M. Merkli: Dynamical localization of quantum walks in random environments, Journal of Statistical Physics, {\bf 140}, 1025-1053 (2010)


\bibitem{segawa}N. Konno, T. Luczak, and E. Segawa: Limit measures of inhomogeneous discrete-time quantum walks in one dimension, Quantum Information Processing {\bf 12}, 33-53 (2013)

\bibitem{kempe} N. Shenvi, J. Kempe, and K. B. Whaley: A quantum random walk search algorithm, Physical Review A, {\bf 67}, 052307 (2003)

\bibitem{kitagawa}T. Kitagawa, M. S. Rudner, E. Berg, and E. Demler: Exploring topological phases with quantum walks, Physical Review A, {\bf 82}, 033429 (2010)

\bibitem{konnoweak}N. Konno: A new type of limit theorems for the one-dimensional quantum random walk, Journal of the Mathematical Society of Japan, {\bf 57}, 1179-1195 (2005)

\bibitem{konnopath}N. Konno: A path integral approach for disordered quantum walks in one dimension, Fluctuation and Noise Letters, {\bf 5}, 529-537 (2005)

\bibitem{konno}N. Konno: Localization of an inhomogeneous discrete-time quantum walk on the line,
Quantum Information Processing, {\bf 9}, 405-418 (2010)

\bibitem{konno}N. Konno and E. Segawa: One-dimensional quantum walks via generating function and the CGMV method, Quantum Information and Computation, {\bf 14}, 1165-1186 (2014)

\bibitem{yoo}N. Konno and H. J. Yoo: Limit theorems for open quantum random walks, Journal of Statistical Physics, {\bf 150},  299-319 (2013)

\bibitem{nayak}A. Nayak and A. Vishwanath: Quantum walk on the line, arXiv:quant-ph/0010117 

\bibitem{katsura}Y. Shikano and H. Katsura: Localization and fractality in inhomogeneous quantum walks
with self-duality, Physical Review E, {\bf 82}, 031122 (2010)

\bibitem{schudo}G. Grimmett, S. Janson, and P. F. Scudo: Weak limits for quantum random walks, Physical Review E, {\bf 69}, 026119 (2004)  

\bibitem{wojcik}A. Wojcik, T. Luczak, P. Kurzynski, A. Grudka, T. Gdala, and M. Bednarska-Bzdega: 
Trapping a particle of a quantum walk on the line, Physical Review A, {\bf 85}, 012329 (2012)\\
\end{thebibliography}
\end{document}